# Mode-Division Multiplexing for Silicon Photonic Network-on-chip

Xinru Wu, Chaoran Huang, Ke Xu, *Member, IEEE*, Chester Shu, *Senior Member, IEEE*, and Hon Ki Tsang, *Senior Member, IEEE*

*Abstract*— Optical interconnect is a potential solution to attain the large bandwidth on-chip communications needed in high performance computers in a low power and low cost manner. Mode-division multiplexing (MDM) is an emerging technology that scales the capacity of a single wavelength carrier by the number of modes in a multimode waveguide, and is attractive as a cost-effective means for high bandwidth density on-chip communications. Advanced modulation formats with high spectral efficiency in MDM networks can further improve the data rates of the optical link. Here, we demonstrate an intra-chip MDM communications link employing advanced modulation formats with two waveguide modes. We demonstrate a compact single wavelength carrier link that is expected to support 2x100 Gb/s mode multiplexed capacity. The network comprised integrated microring modulators at the transmitter, mode multiplexers, multimode waveguide interconnect, mode demultiplexers and integrated germanium on silicon photodetectors. Each of the mode channels achieves 100 Gb/s line rate with 84 Gb/s net payload data rate at 7% overhead for hard-decision forward error correction (HD-FEC) in the OFDM/16-QAM signal transmission.

## I. INTRODUCTION

AS the clock frequency of the microprocessors are limited to the range of 4-5 GHz by the heat generated from the on-chip electrical interconnects [1], modern microprocessors have migrated towards parallel computing with many-core processors (MCPs) and systems-on-chip (SoC). On-chip global interconnects between multiple cores and memories must support data rates approaching 1 TB/s [2] without generating excessive heat. Optical interconnects have appeared as a promising technology that can be used to address the bandwidth bottleneck in on-chip data communications [2], [3]. Integration of photonic circuits on silicon provides a potential solution for low-cost, highly scalable and energy-efficient on-chip data communications.

Wavelength-division multiplexing (WDM) is one of the most mature multiplexing technology used in telecommunications and has been widely explored for on-chip optical interconnects [3-7]. However, WDM has limitations in bandwidth density scalability and requires multiple precise wavelength sources, which are not currently available within the cost scenarios of a single-user computer. Mode-division multiplexing (MDM) offers an additional degree of freedom to scale communication bandwidth by utilizing separate guided modes to carry multiple data channels simultaneously using a single wavelength source. Multimode communication in fiber-optic networks [8] has already been widely demonstrated, but MDM is only now emerging as a possible approach for photonic integrated circuits to increase the bandwidth density in on-chip optical networks.

Mode add-drop multiplexers (MADM) are key functional elements for MDM. Many promising approaches have been reported for MADM including asymmetric directional couplers (ADCs) [9-11], single-mode microring- [12] and multimode microring-based [13] couplers, asymmetric Y-junctions [14-17] and topology optimized mode (de)multiplexer [18]. Most of the functional elements for MDM networks have already been demonstrated. A WDM compatible multimode switch using single-mode microring resonator [19] and a thermal-optic Mach-Zehnder interferometer (MZI) based broadband mode switch [20] and a 4-channel silicon photonic MIMO mode unscrambler using a mesh of cascaded MZIs [21] have been demonstrated. Reference [22] described the implementation of a two-mode ring resonator based add/drop filter which has the potential for channel selective filtering in MDM systems. We previously reported a single wavelength 3 x 28 Gb/s MDM on-chip communication link with integrated microring modulators [23]. A high-bandwidth link with silicon modulators and mode multiplexing [24] and an on-chip WDM compatible MDM interconnection with optional demodulation function have also been demonstrated [25]. A monolithic MDM optical interconnect operating at 20 Gb/s data rate was also recently reported [26].

Advanced modulation formats using direct detection such as discrete multi-tone (DMT) [27] - [29] and orthogonal frequency-division multiplexing (OFDM) [30], [31] have attracted much attention for short-reach on-chip communication and for optical interconnects in data centers.

Manuscript received xxxx; revised xxxx; accepted xxxx. Date of publication xxxx; date of current version xxxx. This work was supported by NSFC/RGC joint research grant N_CUHK404/14.

X. Wu, C. Huang, C. Shu and H. K. Tsang are with the Department of Electronic Engineering, the Chinese University of Hong Kong, Hong Kong (e-mail:xrwu@ee.cuhk.edu.hk;crhuang@ee.cuhk.edu.hk;ctshu@ee.cuhk.edu.hk; hktsang@ee.cuhk.edu.hk).

K. Xu is with the Department of Electronic and Information Engineering, Shenzhen Graduate School, Harbin Institute of Technology, Shenzhen 518055, China (e-mail: kxu@hitsz.edu.cn).

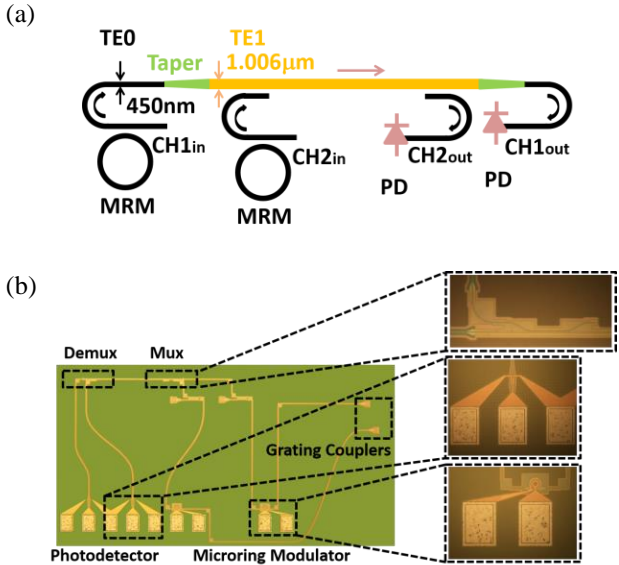

Fig. 1. (a) MDM optical link, including integrated microring modulators, integrated germanium photodetectors and mode add-drop (de)multiplexers. (b) Microscope images of the 2-channel MDM on-chip communications link.

Their high spectral efficiency allows them to achieve 100Gb/s data rate with only 20 GHz optical devices and the simple direct detection receivers offer advantages of lower cost and smaller footprint when compared with optical coherent receivers. Silicon microring modulators [28], [30] and Mach-Zehnder modulators [29] have been reported to attain over 100 Gb/s data rate using OFDM or DMT technologies. In this letter, we present a highly compact and monolithically integrated MDM optical communication link, and show its suitability for use with 16-quadrature amplitude modulation (16-QAM) encoded OFDM direct-detection communications for use in on-chip optical interconnects. The successful experimental demonstration of using different mode-channels (TE0 and TE1 mode channels), each operating at 100 Gb/s line rate, points to the possibility of using many modes in MDM for scaling to TB/s datarates.

## II. NETWORK ARCHITECTURE AND DEVICE DESIGN

In terms of the principal requirements for intra-chip interconnects of low power consumption, high bandwidth and small footprint, microring modulators, with dynamic energy consumption in the order of 10 fJ/bit [32], are one of the most promising candidates for optical modulators for intra-chip optical interconnects. The schematic of the demonstration network reported in this paper using two microring modulators, two germanium photodetectors (Ge PDs) and a MDM link is shown in Fig. 1 (a). In this optical link, a single continuous-wave (CW) laser acts as the light source. The laser is coupled to the waveguide via a shallow-etched focusing grating coupler. Two separate silicon microring modulators encode the electrical OFDM signal onto different optical modes of the waveguide. The modulated signals in the single-mode access waveguides are converted to different high order modes in the multimode bus waveguide using cascaded ADCs. In the ADC, the signal encoded on the fundamental mode of the access waveguide can be evanescently coupled to the higher order mode of the multimode bus waveguide. The detailed design of the cascaded ADCs for the mode add-drop multiplexers is described in reference [10]. The modulated signals carried by the fundamental and first order modes propagate through the multimode bus waveguide and are demultiplexed by another set of cascaded ADCs. The ADCs near the receiver distributes each mode from the bus waveguide into different single-mode drop waveguides, and eventual optical to electrical conversion is performed in the integrated germanium on silicon photodiodes.

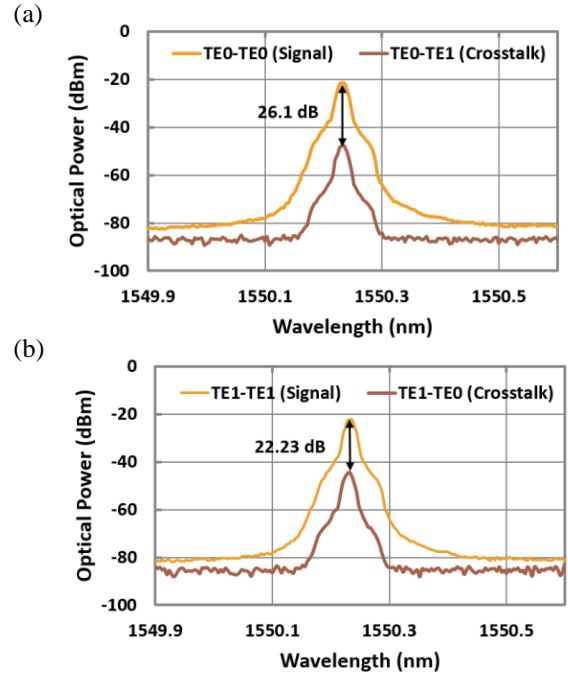

Fig. 2. (a) Transmission and (b) crosstalk measurements for different channels. Desired signal profiles compared with crosstalk profiles measured with 0.02 nm resolution bandwidth. Signal and crosstalk are measured individually at a fixed wavelength of 1550.23 nm.

The MDM monolithically integrated circuits were fabricated on a silicon-on-insulator (SOI) platform with a top silicon thickness of 220 nm. The ADCs for the mode add-drop multiplexers shown in Fig. 1 (a) comprise 450 nm width input and output single-mode waveguides and 1.006 μm width waveguides with 200 nm coupling gap for $TE_1$ mode (de)multiplexers. Both microring modulators are identical, and have a ring radius of 7.5 μm. The nominal coupling gap between ring resonator and bus waveguide is 180 nm and the dimension of silicon rib waveguide is 500 nm × 220 nm. Power coupling from bus waveguide to ring cavity is about 5.5% at 1550 nm, which is designed to be equal to the ring cavity loss, thus to achieve critical coupling. There is a slab waveguide of 60 nm thickness beside the rib waveguide to provide an electrical path for the free carrier modulation. The design of the modulator and the doping level is similar to the device described in the reference [33]. The Ge PDs have a length of about 76 μm. Fig. 1 (b) shows the microscope images of the fabricated device.

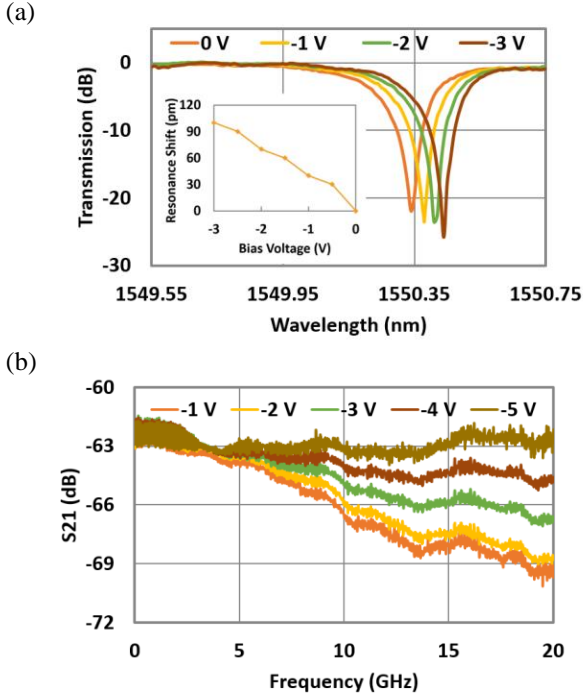

Fig. 3. (a) Transmission spectra of the mciroring modulator at different reverse bias voltages. Inset shows the resonant wavelength shift as a function of the applied bias voltages. (b) Electro-optical response of the microring modulator under different reverse bias. During the measurement, we fix the wavelength at -0.06 nm offset from the resonant wavelength.

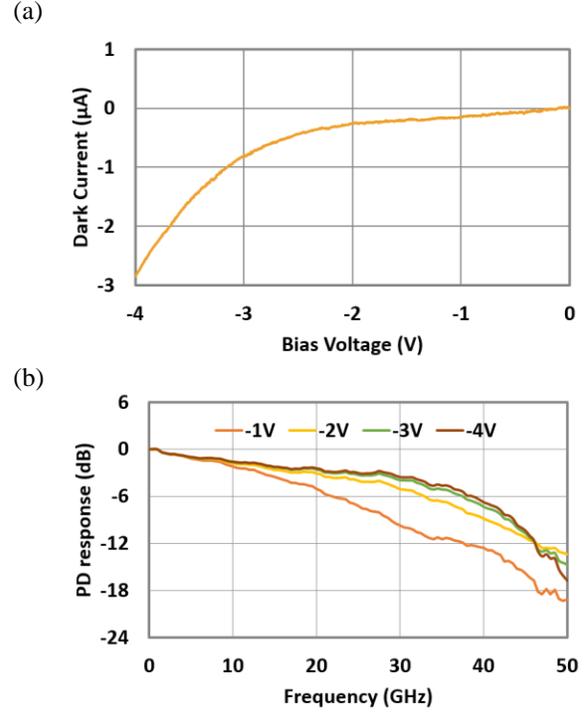

Fig. 4. (a) Measured dark current versus bias voltage of the integrated germanium photodiode. (b) Frequency response of germanium photodiode under different bias voltage, showing 28 GHz 3 dB bandwidth at -4 V bias.

## III. EXPERIMENTAL RESULTS

We first characterize the optical performance of the cascaded ADCs by using a nominally identical test ADC structure on the same chip. Transmission spectra and mode crosstalk are measured by shining a CW laser source with a fixed wavelength, 1550.23 nm, into one channel at a time and measuring the optical spectrum at each of the two output ports one by one. The input grating coupling loss is 3 dB for the TE mode. The output profiles of the desired signal and the crosstalk are then compared as shown in Fig. 2 (a) and (b). The on-chip insertion loss for $TE_0$ mode is 0.5 dB and for $TE_1$ mode is 1.2 dB. The crosstalk between $TE_0$ and $TE_1$ channels is less than -22.2 dB. Mode multiplexer insertion loss is mainly attributed to the waveguide propagation loss and the bi-level taper transition between the shallow-etched waveguide and fully-etched waveguide. The mode crosstalk is increased by the in-complete coupling arising from small deviations of the actual fabricated structure from the design. Keeping a low level of optical crosstalk is important as the optical crosstalk would result in leakage of power to adjacent channels, and thus increase insertion loss and degrade the effective signal to noise ratio.

High-speed modulation for microring modulator is achieved by controlling the depletion width of the PN junction in the microring waveguide, and thus change the waveguide's refractive index and resonant wavelength. The larger the reverse bias, the larger the depletion width and the larger increase in effective index, which results in resonance red shift as shown in Fig. 3 (a). Fig. 3 (a) shows the transmission spectra of the modulator as a function of reverse bias. The extinction ratio is larger than 22 dB at 0V bias and the resonance red shift is 33 pm/V as plotted in the inset of Fig. 3 (a). The measured optical 3 dB bandwidth is 0.23 nm and the estimated loaded quality factor is 6740. We measure the small signal electro-optical response of the modulator in the frequency range of 9 KHz to 20 GHz by using a network analyzer (Keysight E5071C). The measured 3 dB bandwidth of the microring modulator used in this experiment is about 15 GHz at -3 V bias voltage. Under -5 V bias, which indicate large optical detuning, peaking effect is observed at the roll-off of the final frequency response [34].

For direct characterization of the Ge PD without the ADCs insertion losses, we use an identical test Ge PD on the same chip connected directly to a grating coupler. Fig. 4 (a) shows the dark current versus the bias voltage. At 4 V reverse bias, the measured dark current is about 2.9 μA. Fig. 4 (b) shows the frequency response of the Ge PD which is obtained from time domain impulse-response measurement using a femtosecond fiber laser (Toptica FemtoFErb 1560) and a 50GHz sampling oscilloscope (Agilent 86100A). The measured 3 dB bandwidth of the detector is 28GHz at -4 V bias.

## IV. OFDM MODULATION AND DETECTION

High-speed operation of the integrated MDM optical link is tested with OFDM signal modulation for each mode channel individually. Fig. 5 depicts the experimental setup of the OFDM modulation and direct detection system. We set the input wavelength at 1550.23 nm and the incident fiber power at 5 dBm. Light from a tunable laser is passed through a polarization controller (PC) and then launched into each channel separately via the waveguide grating couplers. The electrical OFDM signal is generated by the transmitter-digital signal processor (Tx-DSP) and an arbitrary waveform generator (AWG) (Keysight M8195A) with 25 GHz analog bandwidth and 65 Gsample/s. The OFDM signal consists 256 subcarriers with 16-QAM encoding and 16-point cyclic prefix (CP). The generated OFDM signal from the AWG is amplified to a peak-to-peak voltage ($V_{pp}$) of ~3.5 V. The electrical OFDM signal with a -1.8 V DC bias is then applied to the mciroring modulator through a high-speed RF probe. -1.8 V DC bias was applied to the microring modulator to set it at -4 dB from the maximum transmission. The silicon modulator converts the electrical signal to the optical domain. The modulated signal is then converted to the prescribed higher order mode through the mode multiplexer and propagate through the multimode bus waveguide. At the receiver the multimode signal is demultiplexed to the fundamental mode and directly detected by the integrated Ge PD. The electrical output signal is coupled from the silicon chip by another high-speed RF probe and then captured by a real-time oscilloscope (Teledyne Lecroy LabMaster 10-25Zi) at a capture rate of 80 GSample/s. Since the number of effective bits of the real-time oscilloscope is only 4.7 bits, for better amplitude resolution, a 45 GHz broadband RF amplifier is used to amplify the electrical signal to about 10 dBm before it is recorded by the real-time oscilloscope. The data from the received OFDM signal is demodulated offline by receiver-digital signal processor (Rx-DSP). The Rx-DSP includes resampling, frame synchronization and removal of the CP. Finally, channel equalization, bit error rate counting and constellation demapping are carried out to decode the data.

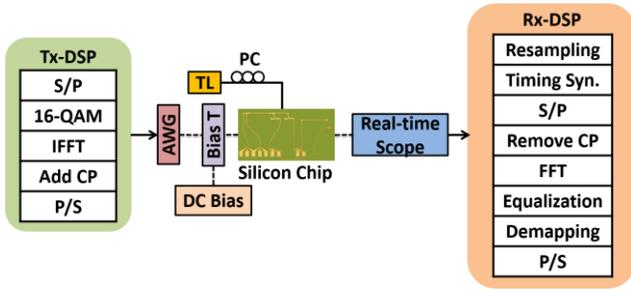

Fig. 5. Experimental setup of high-speed OFDM signal transmission, including tunable laser (TL), polarization controller (PC), arbitrary waveform generator (AWG), transmitter-digital signal processor (Tx-DSP), receiver-digital signal processor (Rx-DSP).

Fig. 6 (a) and (b) depict the signal-to-noise ratio (SNR) of each subcarrier for the OFDM communication link of the two channels. The SNRs drop to 12 dB at 25 GHz for both channels. Fig. 7 (a) and (b) show the measured BERs at different data rate for the two mode channels, respectively. The received optical power for $TE_0$ channel is -2.5 dBm and for $TE_1$ channel is -3.4 dBm. Using 20% coding overhead for hard-decision forward error correction (HD-FEC) [35], which has a BER threshold of $1.5 \times 10^{-2}$, the $TE_0$ channel can achieve a data rate of ~128 Gb/s with > 105 Gb/s net information rate and $TE_0$ channel can achieve ~ 120 Gb/s with ~ 100 Gb/s net information rate. At 7 % overhead HD-FEC, which has a BER threshold of $3.8 \times 10^{-3}$, 100 Gb/s aggregate data rate could be achieved for both channels with 84 Gb/s net information rate. The achievable data rate for $TE_1$ channel is slightly lower than that of $TE_0$ channel because of the on-chip insertion loss for TE1 channel is 0.7 dB larger than that of TE0 channel. Inset of Fig. 6 (a) and (b) show the constellation diagrams of each channel operating at 100 Gb/s line rate.

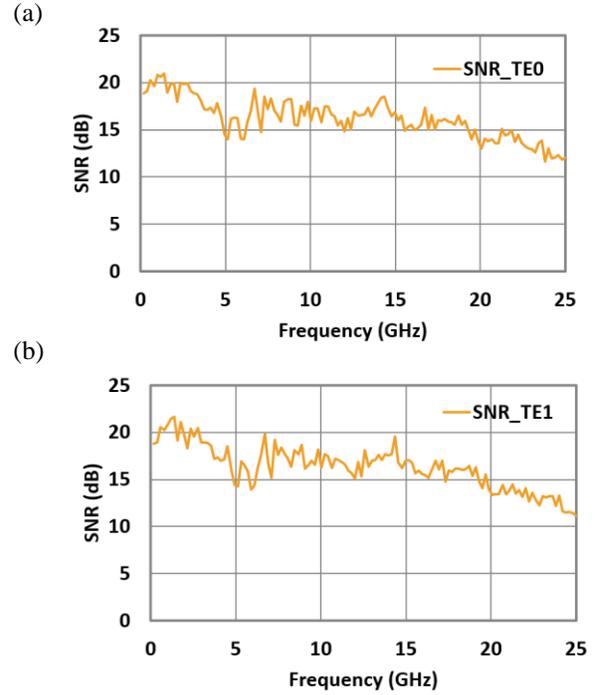

Fig. 6. Measured SNRs of each subcarrier for (a) TE0 channel and (b) TE1 channel, respectively.

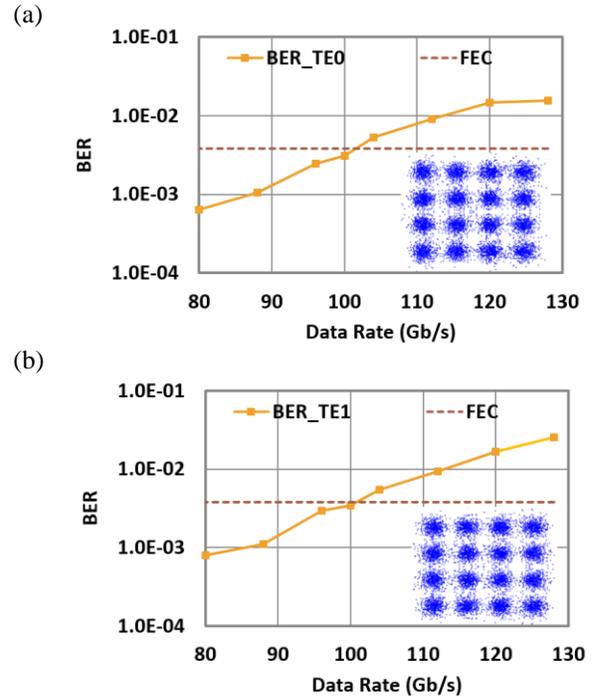

Fig. 7. (a) (b) BER performance of OFDM signal modulation at different data rate for $TE_0$ channel and $TE_1$ channel, respectively. Insets are the constellation diagrams of the received 100 Gb/s line rate signal for each channel.

## V. Discussion and Conclusion

The data transmission measurements demonstrate the potential viability of high-speed MDM optical communications for on chip optical interconnects. Further characterization of the circuits, including a MDM link simultaneously with multiple modes and measurement of channel crosstalk with multimode waveguide bends will be needed. In the test chip fabricated for this paper, we could not couple to the two mode channels simultaneously because the pitch between the two input grating couplers does not match the available multicore fiber array. Therefore, we have estimated the OSNR penalty following the approach in reference [36] using the measured inter-channel crosstalk (Fig. 2). The theoretical model in [36] regards the crosstalk as a virtual additive white Gaussian noise on I-Q planes. The -22.2 dB crosstalk will introduce about 0.925 dB OSNR penalty when BER is $1 \times 10^{-3}$. This agrees well with the crosstalk tolerance of QAM described in reference [37]. Although mode crosstalk is a contributing factor to signal degradation, our results indicate that 2 x 100 Gb/s aggregate data rate should be possible to be realized.

Further performance improvement and data rate increase can be expected by improving the noise performance of the receiver. Since no integrated low-noise transimpedance amplifier (TIA) is used with the integrated Ge PDs, an RF power amplifier designed for modulators with a wider noise bandwidth of 45 GHz and 6 dB noise figure at 15 GHz is used in the above measurements. The electrical noise degrades the received SNR performance. We think MDM may be easily scaled using different higher order waveguide and datarates may be further doubled by using polarization diversity. When an array of microring modulators and Ge PDs are integrated with five-mode cascaded polarization-diversity mode multiplexers, over 1 Tb/s aggregate data rate may be achieved for a single wavelength intra-chip MDM optical link.

In conclusion, we have demonstrated an intra-chip optical communication link by using a monolithically integrated MDM photonic integrated circuit. Insertion loss for the two channels are less than 1.2 dB and the mode crosstalk between two channels are less than -22.2 dB. Aggregate data rate of 100 Gb/s with 84 Gb/s net data rate OFDM/16-QAM signal transmission is achieved for each channel at 7 % overhead HD-FEC. The proposed on-chip MDM system is promising for high density optical interconnects and cost effective optical networks on chip.

Acknowledgement

The authors thank IMEC for device fabrication from the ePIXfab initiative and Prof. Daoxin Dai of Zhejiang University for useful discussions.